\documentclass[page-classic]{ulas_cls} 

\usepackage{amsmath}
\usepackage{graphicx}
\usepackage{amsfonts}
\usepackage[utf8]{inputenc}
\usepackage{hyperref}
\usepackage[english]{babel}
\graphicspath{{fig/}}

\newcommand{\ave}[1]{\left \langle  #1 \right \rangle}
\newcommand{\textpn}[1]{\textbf{\textsf{#1}}}
\newcommand{\unaryminus}{\scalebox{0.5}[1.0]{\( - \)}}
\newcommand{\smallcal}[1]{\scalebox{0.8}{\( \mathcal{#1} \)}}

\title{The configuration multi-edge model: Assessing the effect of fixing node strengths on weighted network magnitudes}

\author{O. Sagarra\inst{1} \and F. Font-Clos\inst{2,3} \and C. J. P\'erez-Vicente\inst{1} \and A.  D\'{\i}az-Guilera\inst{1}}
\shortauthor{O. Sagarra \etal}

\institute{                    
  \inst{1} Departament de F\'{\i}sica Fonamental - Universitat de Barcelona, Facultat de F\'isica C. Mart\'i i Franqu\`es 1 - 08028 Barcelona, Spain\\
  \inst{2} Centre de Recerca Matem\`atica - Campus de Bellaterra, Edifici C - 08193 Bellaterra (Barcelona), Spain\\
  \inst{3} Departament de Matem\`atiques - Universitat Aut\`onoma de Barcelona, Edifici C - 08193 Bellaterra (Barcelona), Spain
}

\abstract{
Complex networks grow subject to structural constraints which affect their measurable properties. Assessing the effect that such constraints impose on their observables is thus a crucial aspect to be taken into account in their analysis. To this end, we examine the effect of fixing the strength sequence in multi-edge networks on several network observables such as degrees, disparity, average neighbor properties and weight distribution using an ensemble approach. We provide a general method to calculate any desired weighted network metric and we show that several features detected in real data could be explained solely by structural constraints. We thus justify the need of analytical null models to be used as basis to assess the relevance of features found in real data represented in weighted network form.
}

\begin{document}

\maketitle

\section{Introduction}\label{sec: intro}
Modern complex networks theory has found many applications since its dawn. In particular, the explosion of information technologies has given rise to large-scale, high-dimensional data sets, in which hidden relations might now be uncovered. This has fostered data-driven studies in a wide spectrum of fields ranging from biology \cite{serrano2009extracting} to social sciences \cite{Stehle2013604}, including transportation studies \cite{Barthelemy2011}, genomics \cite{Uhlmann2012}, ecology or bibliometrics \cite{camacho2002robust}.
The repertoire of available networks for data modeling has thus grown accordingly: binary or weighted \cite{Zlati2010,Newman2004b}, directed or undirected, and also simple or multilayer \cite{Gomez2013a,DeDomenico2013} structures have been used. However, extending some of the most basic concepts and tools, such as clustering coefficient, centrality measures, and even finding suitable null models in each case has proven harder than expected \cite{Saramaki2007}, giving rise to multiple definitions in some cases and, consequently, to some controversy.

In fact, the need to distinguish different types of so-called \textit{weighted} networks according to the nature of the events being represented has been pointed out recently \cite{Sagarra2013c}. If nodes accept multiple \emph{distinguishable} connections, then one can speak of multi-edge networks, and it is in this scenario where we propose a flexible, general theory for null-model generation. Our results allow to compute exact analytical expressions for network observables generated under random conditions but preserving some given properties or constraints from the original data \cite{Squartini2011c,Annibale2009a}.  This allows to quantify the relevance of the observed features, which given the high-dimensionality of the studied data sets in real complex networks might not be trivial to detect otherwise \cite{Bianconi2009a}. A classical example where this circumstance is important are Origin-Destination matrices (OD), where the mobility of agents between departure and arrival nodes is represented as weighted networks: usually the average total flow incoming or outgoing from each site, which corresponds to the strength of a node, is constrained by several factors such as population or density of commercial areas. This fact needs to be assessed on the observed data as it can generate complicated spatial patterns that can produce spatial correlations such as the so-called \textit{gravity laws of transportation} \cite{Barthelemy2011}.

In this letter, the effect of fixing an arbitrary strengths sequence on several network observables is thoroughly studied. The obtained results serve as a complement to its binary counterpart, the classical configurational model for arbitrary degree sequences \cite{Molloy1995a}. We develop the full edge and node statistics as well as first order correlations using an application of a general ensemble approach developed in \cite{Sagarra2013c}, providing not only average expected values for the observables but also precise bounds for its fluctuations and we compare the obtained results with simulations, yielding excellent agreement. By particularizing our general results to the case of power law distributed strengths, commonly found in real data \cite{Colizza2006a,Kaluza2010a,Roth2011,DeMontis2005,Barrat2013a}, we demonstrate how the null-model expectations of some widely used weighted network metrics, which are generally considered a sign of relevant correlations (see \cite{Popovic2012} and references therein), can instead be seen as just a consequence of the particular form of the imposed strength sequence, and hence may not represent any unexpected property of the network under study. 

\section{Theoretical Framework}


We start by considering a multi-edge undirected network with a fixed number of nodes $N$ and hence a total of $L=N(N+1)/2$ possible edges. Each edge has an associated integer weight $t_{ij}\in{\{0,1,2\dots\}}$, and the strength of a node $i$ is defined as $s_i = \sum_j t_{ij}$. We further consider a fixed strength sequence $\{\hat{s}_i\}\,; i\in 1,N$, so that we have a total of $\hat{T}=\sum_i \hat{s}_i$ \emph{distinguishable} events to be randomly allocated among $L$ edges. Note that the difference between distinguishable and non-distinguishable events is crucial, as it happens usually in statistical mechanics, because the resulting statistics of the ensembles one can consider depend on this property (see \cite{Garlaschelli2009z,Sagarra2013c} for an extended discussion). As for the notation, note also that in general we will use $\hat{x}$ to denote the value to which a variable $x$ is being constrained

To clarify the elements of our system, we can specify them for the case of an OD study as an illustrative example: Nodes correspond to different locations, events to individual trips between locations, the strengths of nodes to the fixed amount of travelers departing/entering each location and the weights correspond to the observed flows between locations. In this work we consider the case where self-loops are allowed for simplicity but the methodology can be easily extended to the case where no-self-loops are accepted or even to directed networks.

Under the circumstances described, in analogy to statistical mechanics for classical systems, one can consider the Grand-Canonical (GC) ensemble of graphs (see \cite{Bianconi2009b,Squartini2011a} for different examples of other network ensembles) fulfilling \emph{on average} the proposed constraints, \emph{i.e.} $\{\ave{s_i}=\sum_j \ave{t_{ij}} = \hat{s}_i\}$ (throughout this letter $\langle x \rangle$ will refer to the ensemble average of random variable $x$ whereas $\bar{x}$ will refer to the graph average of variable $x$ over a single realization). This means that both the occupation numbers or weights $t_{ij}$ and the node strengths $s_i = \sum_{j} t_{ij} $ are integer random variables, and that each network belonging to the ensemble corresponds to a different realization of such variables. However, relative fluctuations around the constraints $\ave{s_i} = \hat{s}_i$ vanish in the large event limit \cite{Sagarra2013c}. The GC  ensemble yields independent Poisson statistics for the occupation numbers $t_{ij}$ with mean $\langle t_{ij}\rangle = \beta x_i x_j$, where $\{x_i\}$ can be considered as node-specific \emph{hidden variables} \cite{Goh2001,Caldarelli2002,Boguna2003}. The values of $\{ x_i \}$ can be obtained by solving the $N$ saddle point equations $\hat{s}_i = \beta x_i \sum_{j}  x_j$ that define the constraints of the system, which correspond to fixing the (ensemble) average strength of each node in the network. This set of equations is easily solvable analytically yielding $x_i = \hat{s}_i \quad \forall i $ and $\beta = \hat{T}^{-1}$. In this way the expression $\ave{t_{ij}}=\hat{s}_i \hat{s}_j / \hat{T}$ is reached: the left-hand side is the ensemble average of a random variable, while the right-hand side is a result expressed in term of the constraints. In our case the $\{\hat{s}_i\}$ are the only fixed quantities and hence we must take them as the basic variables from which to derive the rest of weighted network properties: All nodes sharing the same strength value $\hat{s}_i=\hat{s}$ are statistically equivalent, and possess self-averaging properties (likewise all edges connecting nodes with the same pair of strength values). 
In what follows, we show how to proceed to obtain some particular network metrics, although the procedure is fully general and permits to obtain any desired property.


\begin{figure}[htbp]
\begin{center}
\includegraphics[width=0.95\columnwidth]{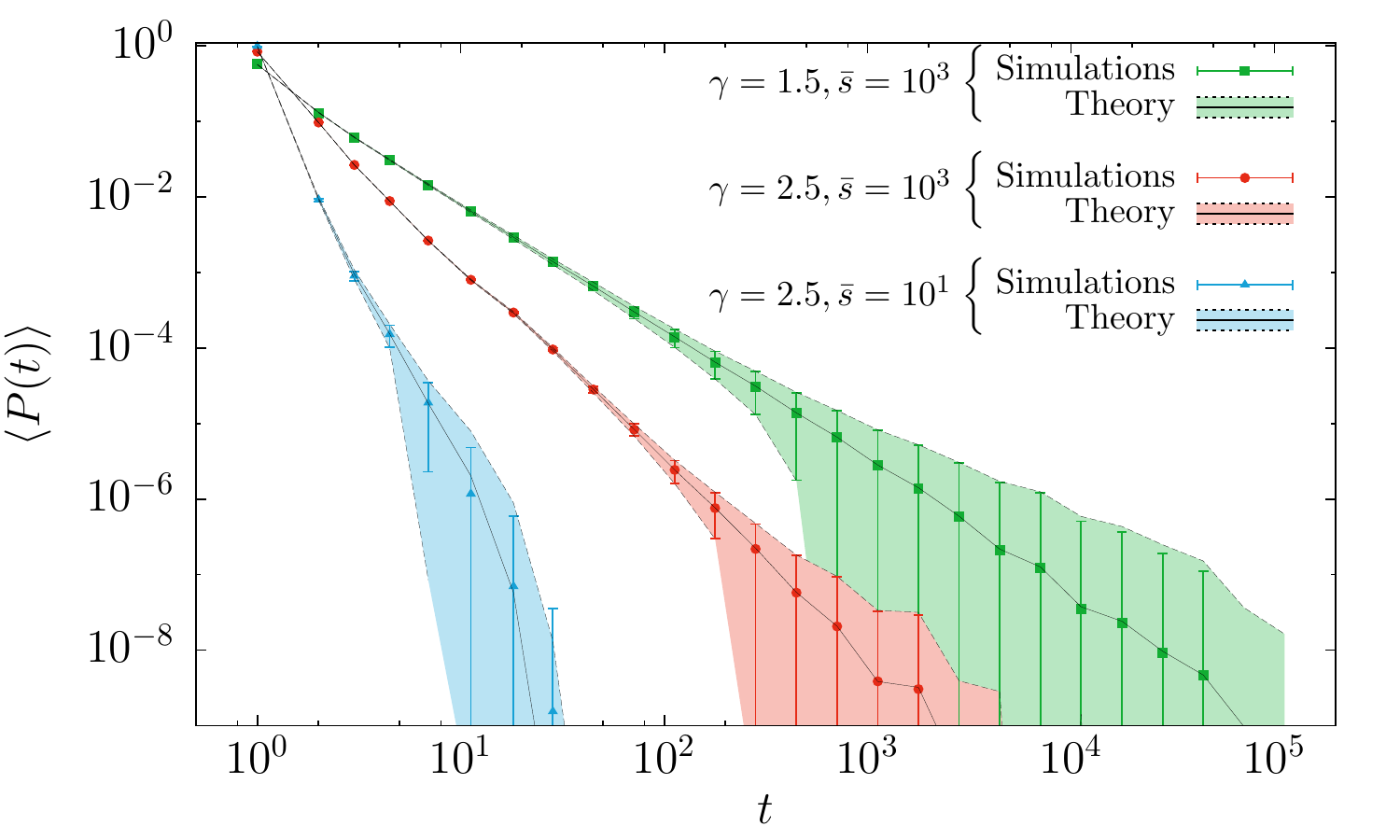}
\includegraphics[width=0.95\columnwidth]{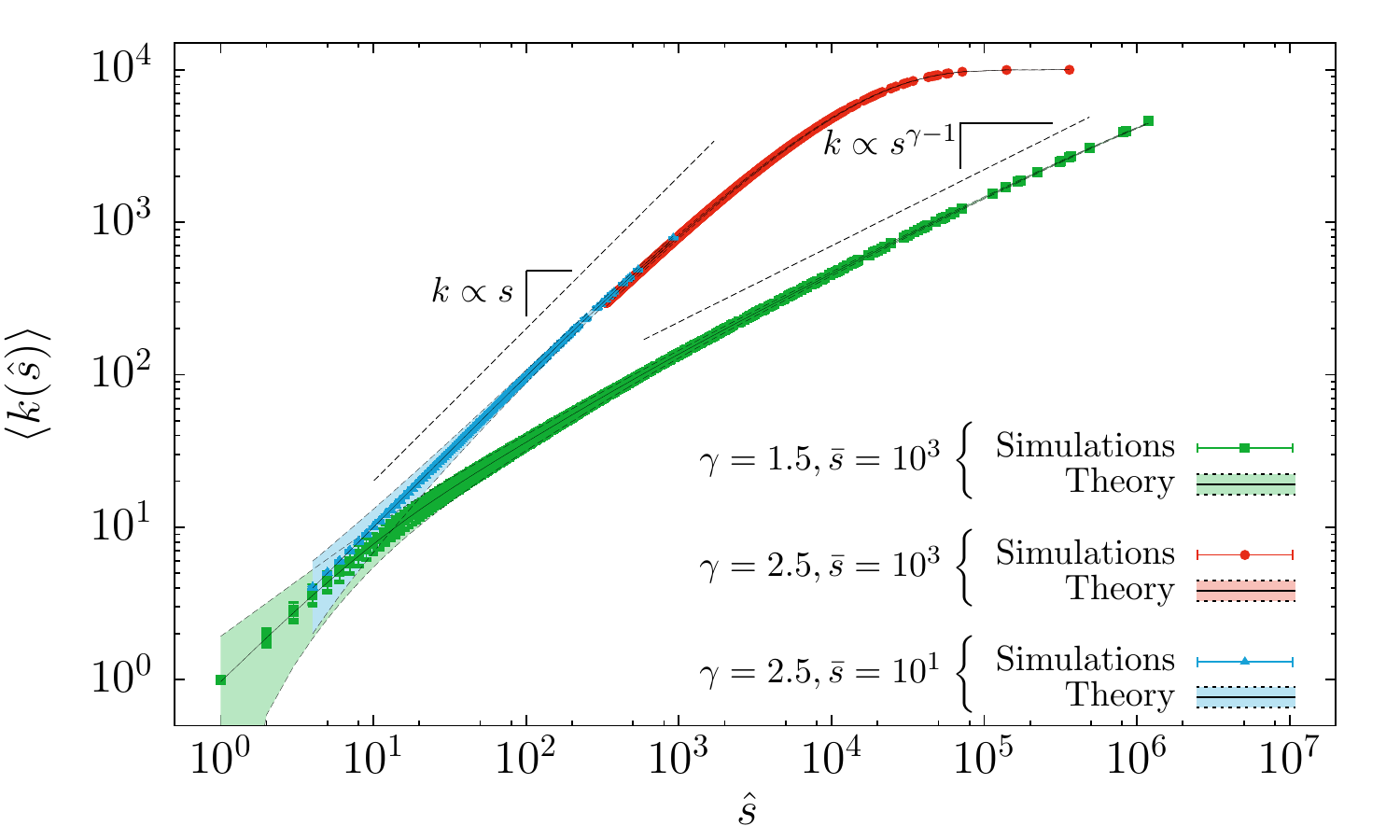}
\caption{(color online) The effect of the strength sequence on network observables. \textbf{(Upper):} Ensemble average of distribution of occupation numbers over existing edges (log-binned) and analytical predictions given by expression \eqref{eq:occnumpt} and its standard deviation (see Appendix, Eq.~\eqref{eq:poft sigma}) for power law distributed strength sequences with $N=10^4$ and different exponents for 1000 repetitions. The dependence on sampling $\bar{s} = \hat{T}/N$ is apparent. \textbf{(Lower):} Degree-strength relationship for the same networks as earlier (average) and theoretical predictions from equation \eqref{eq:ks}. Standard deviation are represented as error bars and lines of constant slope are provided as a guide to the eye.}
\label{fig:occnums_ks}
\end{center}
\end{figure}
\section{General methodology}
Many network metrics widely used in the literature can be written as a quotient of functions of the edge weights $\mathcal{M}=x(t_{ij})/y(t_{ij})$. In our framework, $\{t_{ij}\}$ are random variables and computing $\ave{\mathcal{M}}$ is not straightforward. We thus need to rely on approximations, expanding the expressions in Taylor series around their mean values and then taking the ensemble average of the first terms of the sum.
\begin{alignat}{1}
\label{eq:taylor}
\left \langle \mathcal{M} \right \rangle &\simeq
 \frac{\langle x \rangle}{\langle y \rangle}\left(1+\frac{\langle y^{2}\rangle}{\langle y \rangle^{2}} - \frac{\langle xy \rangle}{\langle x \rangle \langle y \rangle} \right)
 \quad \quad \quad \\
\label{eq:taylor2}
\sigma^2_{\mathcal{M}} &\simeq
\frac{\left \langle x \right \rangle^2}{\left \langle y \right \rangle^2}
\left(
\frac{\left \langle x^2 \right \rangle} { \langle x \rangle ^2}
+\frac{\left \langle y^2 \right \rangle} { \langle y \rangle ^2}
- 2 \frac{\langle xy \rangle}{\langle x \rangle \langle y \rangle}
\right).
\end{alignat}
These expressions can be used to compute expected values and and fluctuations of any network metric expressed as a ratio of functions of the occupation numbers $x(t_{ij}), y(t_{ij})$
provided that 
the moments ($\ave{x},\ave{x^2},\ave{y},\ave{y^2}$) in the right-hand side can be evaluated. 
This is usually the case when $x(t_{ij}),y(t_{ij})$ are algebraic expressions of $\{t_{ij}\}$ (which are uncorrelated random variables).
 For most metrics $\mathcal{M}$ considered in this Letter, the calculations of the moments of $x(t_{ij})$ and $y(t_{ij})$ are lengthly, but follow from a general methodology without further difficulty\footnote{And can be easily implemented using any standard mathematical symbolic software.}. We thus subsume them in the Appendix also explaining the general strategy used in their calculation, stating here only the key results.

\section{Distribution of weights}\label{sec_weights}
We start by computing the distribution of occupation numbers or \emph{weights}
\begin{equation}
\label{eq:p_of_t}
P(t)= \frac{1}{E}\sum_{ij}\delta(t,t_{ij})
\end{equation}
($E=\sum_{ij} \Theta(t_{ij})$ refers to the total number of existing edges), which has been reported to have broad forms on empirical data for airport flow \cite{Colizza2006a}, cargo ship transport \cite{Kaluza2010a}, public transport in cities \cite{Roth2011}, commuting \cite{DeMontis2005} or face to face interactions \cite{Barrat2013a} among others.

Applying Eq.~\eqref{eq:taylor} to the case of $P(t)$ yields,
\begin{equation}\label{eq:occnumpt}
\begin{split}
\langle P(t) \rangle &= \left \langle  \frac{\sum_{ij} \delta(t,t_{ij})}{ \sum_{ij} \Theta(t_{ij}) } \right \rangle \simeq \frac{\sum_{ij} e^{-\ave{t_{ij}} } \ave{t_{ij}}^t}{t! \langle E \rangle} + \mathcal{O}(\langle E \rangle^{-2})\end{split},
\end{equation}
being $\Theta(x)$ the Heaviside step function. Figure \ref{fig:occnums_ks} shows the distribution of occupation numbers for existing edges and its associated standard deviation (see Eq.~\eqref{eq:poft sigma} in the appendix) for three networks generated using power law distributed strength sequences ($\gamma=1.5, 2.5$) and different graph-average strength $\bar{s} = \hat{T}/N$. We can see that the form of the resulting distribution is broad due to the imposed form of the strength sequence, and hence is not a sign \textit{per se} of any interesting property of the multi-edge network being studied. Moreover, its shape strongly depends on the total number of observed events $\hat{T}$.

\section{Degrees and strengths}
Having tackled the occupation number statistics of the network, in what follows we consider its node-related properties. We have that the strengths $s_i=\sum_j t_{ij}$ will also be Poisson distributed random variables, being sums of independent occupation numbers. Moreover, since the binary projection of occupation numbers $\Theta(t_{ij})$ are Bernoulli distributed variables with parameter
$P(t_{ij} > 0 ) =
1- e^{-\langle t_{ij} \rangle}
$ \cite{Sagarra2013c}
one can can also compute the associated degrees $k_i$ of the nodes, which will be sums of independent Bernoulli random variables \footnote{The distribution of such variables is called \textit{Poisson Bernoulli} and has well-studied properties \cite{Serfling1978}, albeit their moments are difficult to compute. One can, however, give bounds to the error committed whenever assuming a Poisson approximation also for the degrees.}. We have that,
\begin{align}\label{eq:ks}
	\begin{split}
		\langle k(\hat{s}_i) \rangle &= \left \langle \sum_{j} \Theta(t_{ij}) \right \rangle = \sum_{j} P(t_{ij} > 0) =\\
		&=\sum_{j} \left (1- e^{-\ave{t_{ij}}} \right) = N- \sum_{j} e^{-\frac{\hat{s}_i \hat{s}_j}{\hat{T}}}
	\end{split}
\\
	\begin{split}
		\sigma^2_{k(\hat{s}_i)} &=\sum_{j} \sigma^2_{\Theta(t_{ij})} = \sum_{j} e^{-\ave{t_{ij}}} \left (1-e^{-\ave{t_{ij}}}\right) =\\
		&= N-\langle k(s_i) \rangle - \sum_{j} e^{-2\frac{\hat{s}_i \hat{s}_j}{\hat{T}}}.
	\end{split}
\end{align}
which constitutes an extremely accurate prediction (see Fig. \ref{fig:occnums_ks} lower panel). The asymptotic cases for the ensemble averages are easy to asses: For small strengths we have that $\hat{s} \ll \hat{T}/\hat{s}' \, \forall \hat{s}'$ which leads expression \eqref{eq:ks} to $\langle k(\hat{s}) \rangle \sim \hat{s}$ (converging to a Poisson distribution for degrees due to the properties of the Poisson Bernoulli distribution), while for large strengths one has $\hat{s} \hat{s}' \gg \hat{T} \, \forall \hat{s}'$ which leads to fully connected nodes $\langle k(\hat{s}) \rangle \sim N$ with vanishing variance. 

Results comparing simulations and equation \eqref{eq:ks} are shown in Fig. \ref{fig:occnums_ks} (lower panel), where an interesting transition is observed for $\gamma<2$: The degrees are exactly equal to the strengths for small values of $\hat{s}$ (as expected by conservation of the edges) and evolve to a scaling of the type $k(\hat{s})\sim \hat{s}^{\gamma-1}$ that finally leads to a saturation due to the bounded nature of the observables ($k(\hat{s})\leq N$).


 The important result to take home here is that \textit{a scaling relation of the kind $k(s) \sim s^\beta$ is not always a reliable trace of correlations.} More concretely, we have seen that in our framework, and for the case of power-law distributed strength sequences in particular, it is solely a consequence of the imposed constraints. In other cases, it might or might not be an indicator of correlations, but one cannot assume either case \emph{a priori}: Since this metric heavilly depends on the strength sequence, it always requires comparison with a null model.


\section{Disparity, Average neighbor properties and general metrics}\label{sec_others}
\begin{figure}[htbp]
\begin{center}
\includegraphics[width=0.95\columnwidth]{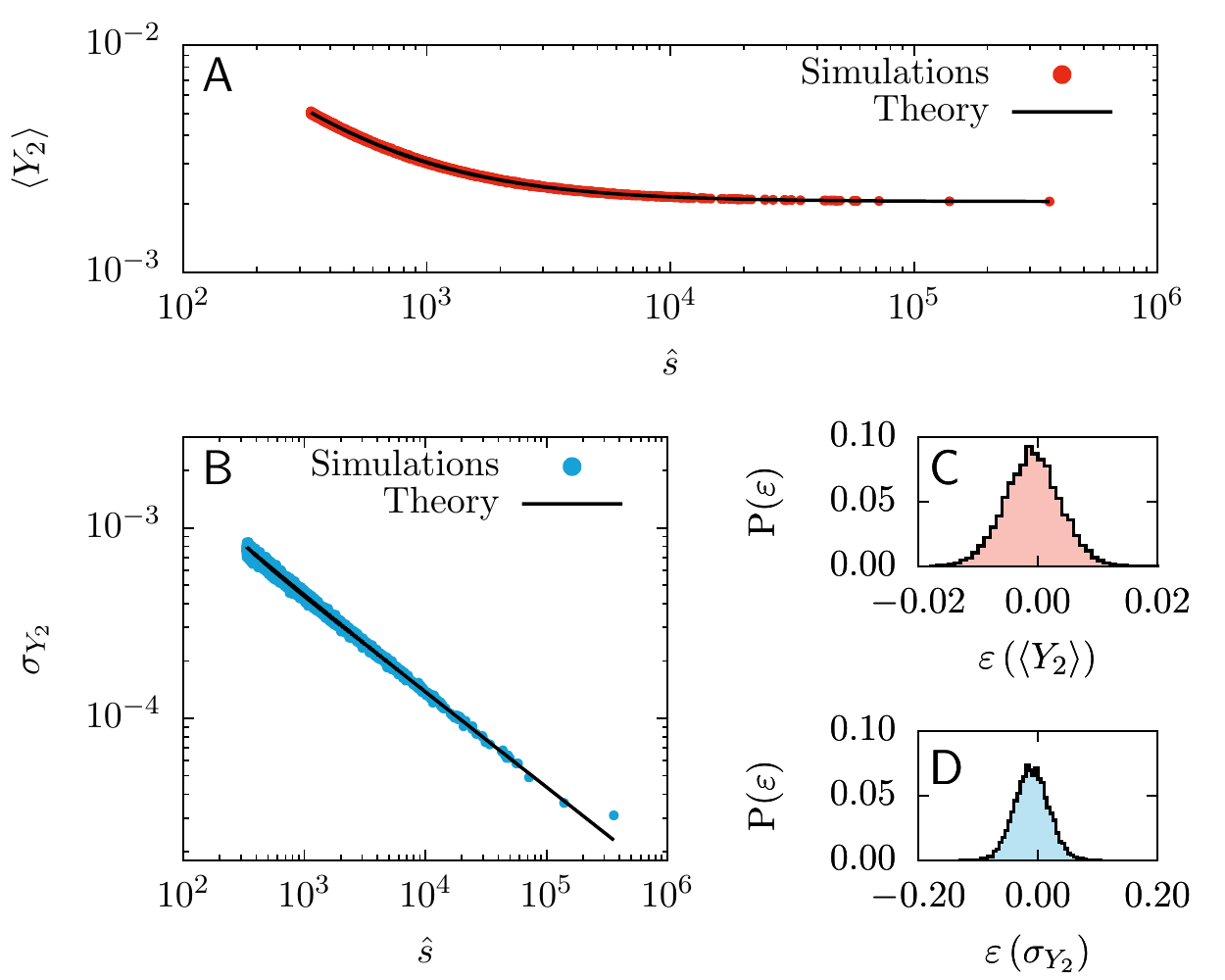}
\includegraphics[width=0.95\columnwidth]{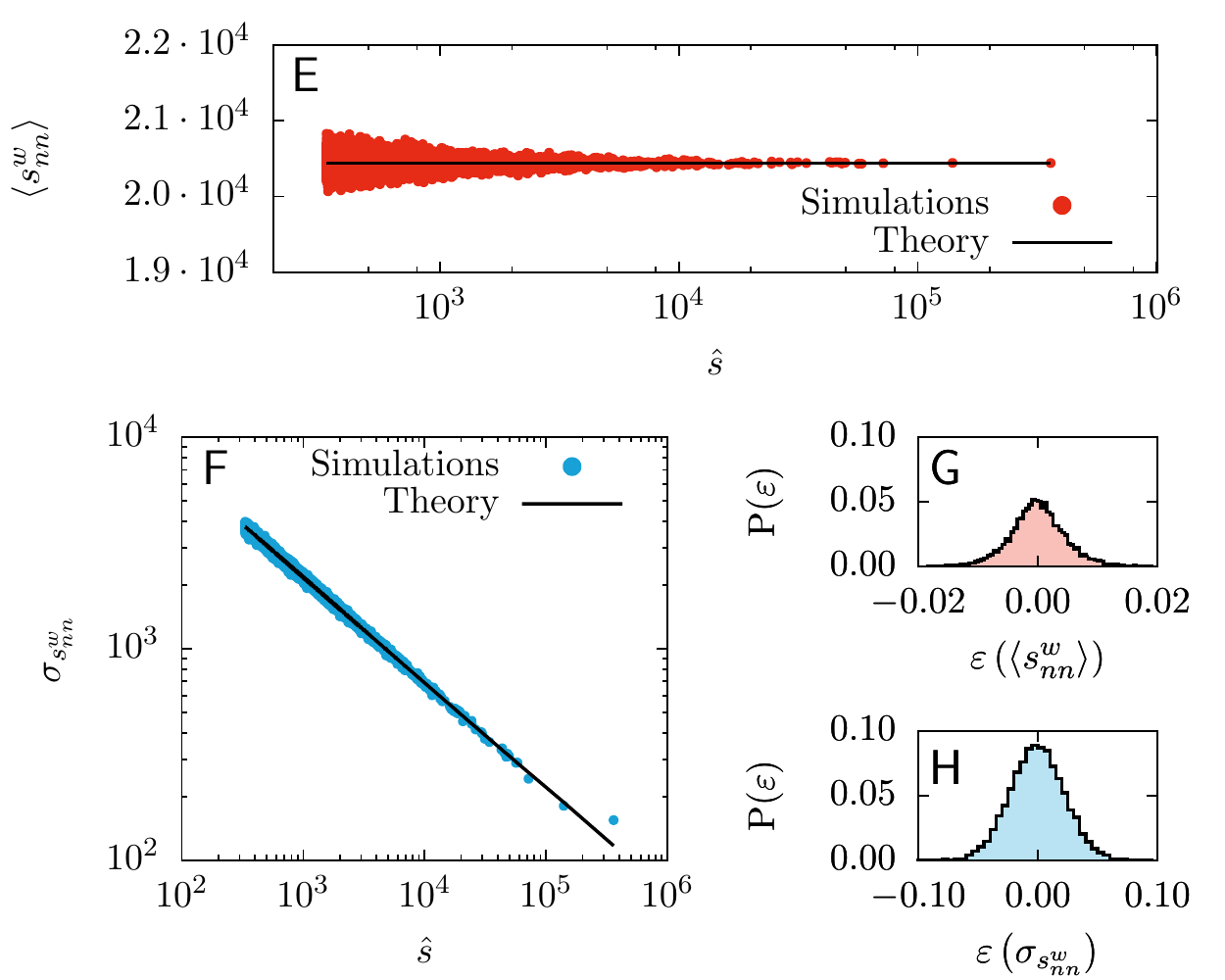}
\caption{(color online) The accuracy of the GC predictions. Ensemble average (\textpn{A},\textpn{E}) and standard deviation (\textpn{B},\textpn{F}) for individual node disparity $Y_2$ and weighted neighbor average strength $s^w_{nn}$  for power law distributed sequences with $\bar{s}=1000$ and $\gamma=2.5$. \textpn{C},\textpn{D},\textpn{G},\textpn{H}: Histogram of relative error between theory and simulations averaged over $1000$ repetitions for the values in Figs. \textpn{A},\textpn{E} and \textpn{B},\textpn{F}. A single outlier corresponding to the lowest value of $\sigma$ for both the disparity and the average neighbor strength is not shown in the histogram on Figs. \textpn{C},\textpn{G}.}
\label{fig:corrs}
\end{center}
\end{figure}
In recent times, efforts have been devoted to extend well-known magnitudes on binary graphs to weighted graphs: Having appropriate null-models for multi-edge graphs permits to assess the applicability of such \textit{weighted} extensions \cite{Ahnert2007}. To this end, one can use the results of the GC ensemble to compute with high accuracy any network metric expressed in terms of $t_{ij}$: As an example we consider widely used magnitudes such as the disparity $Y_2(s_i) = \sum_{i\neq j} t^2_{ij}/s^2_i$  \cite{Serrano2006} and weighted neighbor average strength $s^w_{nn}(s_i) = \sum_j t_{ij} s_j/s_i  $ \cite{Barrat2004b}.
Using again Eq.~\eqref{eq:taylor}, and the fact that $\{t_{ij}\}$ are a set of Poisson-distributed, \emph{independent} random variables, one reaches after some algebra the following expressions
\begin{align}
\langle Y_2(\hat{s_i}) \rangle \simeq&
\frac{1+\frac{\hat{T}_2}{\hat{T}^2} \hat{s}_i}{1+\hat{s}_i}
\left(1+
\frac{\left(\hat{T}^2-\hat{T}_2\right) \left(2
   \hat{s}_i+3\right)}{\left(\hat{s}_i+1\right)^2 \left(\hat{T}_2
   \hat{s}_i+\hat{T}^2\right)}
\right) \label{eq:y2_pred}\\
\langle s^w_{nn}(\hat{s}_i) \rangle &\simeq \left(1+\frac{\hat{T}_2}{\hat{T}}\right)\left(
   1-\frac{\hat{T}_2+\hat{T}\hat{s}_i- \hat{s}_i^2}{\hat{T}(\hat{T}+\hat{T}_2)}
\right). \label{eq:swnn_pred}
\end{align}
where $\hat{T}_n \equiv \sum \hat{s}_i^n$. The average values and their fluctuations are in excellent agreement with the simulations, as can be seen from Fig. \ref{fig:corrs} panels \textpn{A},\textpn{B},\textpn{E},\textpn{F}. The expressions corresponding to $\sigma_{s_{nn}^w}^2$ and $\sigma_{Y_2}^2$, and some details on how to compute them, can be found on the Appendix.

The results show several interesting features: On the one hand, the expectation for the disparity is not $Y_2(\hat{s}) \sim k_i^{-1}$ as assumed under a total random allocation of edge weights \cite{Serrano2006}, but rather decays as $Y_2 \sim \hat{s}_i^{-1}$ and rapidly converges to a plateau, independent of the chosen strength distribution. The weighted average neighbor strength displays an almost flat behavior which is a correct indicator of absence of correlations at the node level. On the other hand, the fluctuations of both magnitudes decay in a power law form as the strength of the node increases: This fact can be easily understood as increased connectedness imply higher availability of sampling.

\section{Simulations}\label{sec_simus}
\begin{figure}[htb]
\begin{center}
\includegraphics[width=0.95\columnwidth]{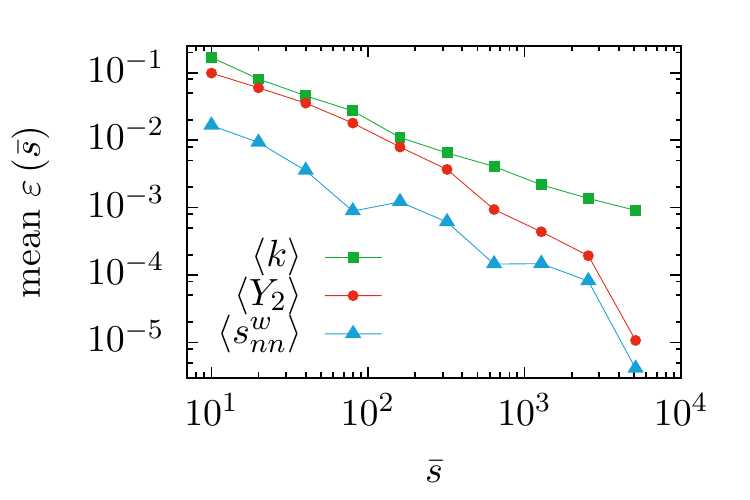}
\caption{(color online) Convergence between ensembles with increased sampling. Relative error between ensemble average predictions and simulations, averaged over all nodes for degree, disparity and average neighbor strength for different values of sampling $\bar{s} = \hat{T} / N$ for 1000 repetitions each point, $\gamma=2.5$ and $N=2000$.}
\label{fig:scaling}
\end{center}
\end{figure}

To quantify the precision of our predictions, we computed the histograms of the relative error generated per node when calculating a given property $z$,  $\varepsilon(z) = (\langle z \rangle_{\mathrm{si}} - \langle z \rangle_{\mathrm{th}})/\langle z \rangle_{\mathrm{si}}$, where the subindices $si$ stand for the Micro-Canonical (MC) simulations\footnote{
The simulations have been performed using a general configuration model rewiring schema with allowed self-loops and multiple collections between links \cite{Serrano2005a}. The code for the generation of multi-edge networks and details on the implementation can be found in \cite{Sagarra2014a}
.} and $th$ for the (GC) theoretical predictions in equations \eqref{eq:y2_pred} and \eqref{eq:swnn_pred}. The histograms in Fig. \ref{fig:corrs} panels \textpn{C},\textpn{D},\textpn{G},\textpn{H} show the accuracy of the obtained results, providing numerical evidence for the equivalence between the MC simulations and the GC predictions, which is expected in the thermodynamic limit when an infinite sampling of events $\hat{T} \to \infty$ is available. Even when this requisite is not met, the use of the theory presented here constitutes an excellent approximation as shown in Fig. \ref{fig:scaling}, where the relative error averaged over all nodes between ensemble expected GC magnitudes and simulations is shown for the different metrics considered 
for a wide range of values of sampling.
\section{Conclusions}\label{sec_concl}
In the present letter, we have provided a theoretical framework for multi-edge networks with a fixed strength sequence that can be used to assess the impact of this structural constraint on the observables of real networks. We provide closed statistical forms for the edge and node statistics by means of a GC ensemble formalism, which coincide with great accuracy with simulations in the Micro-Canonical ensemble, for which we additionally provide the software in \cite{Sagarra2014a}.

The general results for any given strength sequence have been studied, and precise analytical closed forms of both the average expected values and fluctuations of some widely used network metrics such as the existing occupation number distribution, node degree and strength, disparity and first order neighbor correlations have been obtained. We have further used the case of power law distributed sequences as an example to show that the effect of the skewness of this distribution over some widely used network metrics can explain some of the correlations detected in real weighted complex networks.

The effects of the strength sequences are remarkable in all the metrics considered, which stresses the need for real data to be tested against appropriate null models to assess the relevance of observed properties. Interestingly,  second order neighbor correlation properties such as clustering coefficients and its different weighted versions \cite{Saramaki2007} could be similarly computed with the methodology presented, albeit they lead to more complicated expressions. We leave this to future research. 

The blessing of big data may also be its dearest danger: High dimensionality data sets require sophisticated null-models to detect the effects of the system constraints on the given observables and hence comparison with a null model to assess the relevance of observed features in real data is always needed. The present application of ensemble theory to networks yielding exact results for both average values and fluctuations aims to draw attention to this problem and to close this gap for the case of multi-edge networks.

\acknowledgments
We thank M. Szell for comments and suggestions. This work has been partially supported by the EU-LASAGNE Project, Contract
No. 318132 (STREP), the Spanish MINECO Grants FIS2012-38266-C02-02 and FIS2012-31324 and by the Generalitat de Catalunya 2009-SGR-00838. O.S. and F.F have been supported by the Generalitat de Catalunya through the FI Program.

\section{Appendix}
We gather here the expressions for the standard deviation of all the metrics considered. 

\begin{alignat}{1}
\label{eq:poft sigma}
\sigma^2_{P(t)} &\simeq
\frac{\smallcal{Q}_1}{\langle E \rangle^{2}}\! \left(
\!	1 \! - \! \frac{\smallcal{Q}_2}{\smallcal{Q}_1 } 
 + \frac{1}{\langle E  \rangle \smallcal{Q}_1}\left[ 
 1 
\! -\! 2 \frac{\smallcal{R}_1}{\smallcal{Q}_1}
 \right]
 \! -  \! \frac{\smallcal{R}_2}{\langle E \rangle^{2} \smallcal{Q}_1}
\right)
\\
\label{eq:y2_sigma}
\sigma_{Y_2}^2 &\simeq  \hat{T}_1^{-2}
\left(
	\frac{a_3 \hat{s}_i^3 + a_2 \hat{s}_i^2 + a_1 \hat{s}_i + a_0}{\left(\hat{s}_i+1\right)^4}
\right)\\
\sigma^2_{s_{nn}^w} &\simeq
b_{\unaryminus 1} \frac{1}{\hat{s}_i}
+b_{0}
+b_2 \hat{s}_i^2
+b_3 \hat{s}_i^3.
\end{alignat}

where $E$ is the number of present edges and
we have defined the following notation:

\begin{alignat}{2}
p_{ij}(t)&\equiv e^{-\ave{t_{ij}} } \ave{t_{ij}}^t/t!
&
\ave{t_{ij}}&=\hat{s}_i \hat{s}_j / \hat{T} 
\\
\smallcal{Q}_1&=\sum_{ij}p_{ij}(t)
&
\smallcal{R}_1&=\sum_{ij}p_{ij}(t)p_{ij}(0)
\\
\smallcal{Q}_2&=\sum_{ij}(p_{ij}(t))^2
&
\smallcal{R}_2&=\sum_{ij}(1-p_{ij}(0))^2
\\
a_3 &= -4 \left(\hat{T}_2^2-\hat{T}_1 \hat{T}_3\right)
&
b_2 &= 2 \hat{T}_2 / \hat{T}_1^3
\\
a_0 &= 2 \hat{T}_1^4-2 \hat{T}_2 \hat{T}_1^2
&
b_3 &= -2 / \hat{T}_1^2
\end{alignat}
\begin{alignat}{1}
a_2 &= 2 \left(\left[\hat{T}_1^2-5 \hat{T}_2\right] \hat{T}_2+4 \hat{T}_1 \hat{T}_3\right)\\
a_1 &= \hat{T}_1^4+2 \hat{T}_2 \hat{T}_1^2+4\hat{T}_3 \hat{T}_1-7 \hat{T}_2^2\\
b_{\unaryminus 1} &=\left( \hat{T}_1\left[\hat{T}_2+\hat{T}_3\right]-\hat{T}_2^2 \right) \hat{T}_1^{-2}\\
b_0 &= \left(-\hat{T}_2^2+\hat{T}_1 \left[\hat{T}_2+3 \hat{T}_3\right]-\hat{T}_4\right) \hat{T}_1^{-3}
\end{alignat}
with $\hat{T}_n \equiv \sum_i \hat{s}_i^n$.
The calculations leading to these results are admitedly tedious, but follow directly from Eqs.~\eqref{eq:taylor} and \eqref{eq:taylor2} and are of no particular difficulty beyond algebraic manipulation and carefull reordering of the sums. As an illustrative example, consider the disparity $Y_2(s_i)$ for node $i$, defined as
\begin{equation}
Y_2(s_i) = \sum_{j \neq i}\frac{ t^2_{ij}}{s^2_i}
\end{equation}
Identifying $x\equiv \sum_{j \neq i} t^2_{ij}$ and $y \equiv s_i^2 $, Eq.~\eqref{eq:taylor} can be readily applied. In order to approximate $\ave{Y_2(s_i)}$ and $\sigma_{Y_2}$, we need to compute $\ave{x}, \ave{y}, \ave{x^2}, \ave{y^2}$ and $\ave{xy}$. Let us show in full detail, as an illustrative example,  how to compute $\ave{x^2}$ in this case. First, we expand $x^2$ as follows,
\begin{equation}
{x^2} ={\sum_{
\begin{subarray}{l}
j \neq i,\\  k \neq i
\end{subarray}
} t_{ij}^2 t_{ik}^2} = 
	 \sum_{
\begin{subarray}{l}
j \neq i,\\  k \neq j,i
\end{subarray}
}t_{ij}^2 t_{ik}^2+ \sum_{j \neq i} \left[ t_{ij}^4 + 2 t_{ij}^2 t_{ii}^2 \right] + t_{ii}^4,
\end{equation}
so that when the ensemble average is taken, all products factorize (they correspond to different pairs of values $i j$, which are independent),
\begin{equation}
\ave{x^2} = \sum_{
\begin{subarray}{l}
j \neq i,\\  k \neq j,i
\end{subarray}
} \ave{t_{ij}^2} \ave{t_{ik}^2} + \sum_{j \neq i} \left[ \ave{t_{ij}^4} + 2\ave{t_{ij}^2}\ave{t_{ii}^2}  \right] + \ave{t_{ii}^4}
\end{equation}
Finally, since the variables $t_{ij}$ are Poisson-distributed, we can compute their moments ($\ave{t_{ij}^2}=\ave{t_{ij}}(1+\ave{t_{ij}})$), and using that $\ave{t_{ij}}=\hat{s}_i \hat{s}_j / \hat{T}$, and after some algebra, we get to
\begin{equation}
\label{y2x2}
\ave{x^2} = \frac{\hat{T}_2^2 \hat{s}_i^4}{\hat{T}_1^4}+\left(\frac{2 \hat{T}_2}{\hat{T}_1^2}+\frac{4
   \hat{T}_3}{\hat{T}_1^3}\right) \hat{s}_i^3+\left(\frac{6 \hat{T}_2}{\hat{T}_1^2}+1\right) \hat{s}_i^2+\hat{s}_i 
\end{equation}
The rest of the terms can be computed in a similar vein, leading to
\begin{alignat}{1}
	\label{y2x}
   \ave{x} &= \frac{\hat{T}_2 \hat{s}_i^2}{\hat{T}_1^2}+\hat{s}_i \qquad
   \ave{y} =\hat{s}_i^2 + \hat{s}_i \\
   \ave{xy} &= \frac{\hat{T}_2 \hat{s}_i^4}{\hat{T}_1^2}+\left(\frac{5 \hat{T}_2}{\hat{T}_1^2}+1\right) \hat{s}_i^3+\left(\frac{4 \hat{T}_2}{\hat{T}_1^2}+3\right) \hat{s}_i^2+\hat{s}_i \\
\label{y2y2}
   \ave{y^2} &=  \hat{s}_i^4+6 \hat{s}_i^3+7 \hat{s}_i^2+\hat{s}_i
\end{alignat}
Finally, inserting Eqs.~(\ref{y2x2}-\ref{y2y2}) into Eq.~\eqref{eq:taylor} and some simplification leads to the desired result, Eqs.~\eqref{eq:y2_pred}~and~\eqref{eq:y2_sigma}.
\bibliographystyle{eplbib}
\bibliography{new_lib}

\end{document}